\pdfoutput=1
\documentclass[letterpaper,twocolumn,showpacs,floatfix]{revtex4}

\RequirePackage{graphicx,amsmath,amssymb,bm}
\RequirePackage{hyperref}
\hypersetup{%
  breaklinks = {true},
  citecolor = {blue},
  colorlinks = {true},
  linkcolor = {red},
  pdfauthor = {\textcopyright\ Dario A. Bahamon},
  pdfcreator = {\LaTeX\ and \flqq hyperref\frqq},
}

\newcommand{\Fref}[1]{Fig.~\ref{#1}}
\newcommand{\Eqref}[1]{Eq.~(\ref{#1})}
\renewcommand{\eqref}[1]{eq.~(\ref{#1})}

\begin{document}

\title{Effective contact model for geometry-independent conductance 
calculations in graphene}

\author{D. A. Bahamon}
\author{A. H. Castro Neto}
\author{Vitor M. Pereira}
\affiliation{Graphene Research Centre \& Department of Physics,
National University of Singapore, 2 Science Drive 3, Singapore 117542}

\date{\today}

\begin{abstract}
A geometry-independent effective model for the contact self-energies is
proposed to calculate the quantum conductance of patterned graphene devices
using Green's functions. A Corbino disk, being the simplest device where the 
contacts can not be modeled as semi-infinite ribbons, is chosen to illustrate 
this approach. This system's symmetry allows an analytical solution against 
which numerical calculations on the lattice can be benchmarked. The effective 
model perfectly describes the conductance of Corbino disks at low-to-moderate 
energies, and is robust against the size of the annular device region, the 
number of atoms on the edge,  external magnetic fields, or electronic disorder. 
The contact model considered here affords an expedite, flexible, and 
geometry-agnostic approach easily allows the consideration of device dimensions 
encompassing several million atoms, and realistic radial dimensions of a few 
hundreds of nanometers. 

\end{abstract}



\maketitle

To understand the transport properties and predict the performance of 
nano-devices it is necessary to fabricate  appropriate contacts \cite{Hipps}. 
Their reduced size and different geometries have led to successive 
reevaluations of existing techniques and processes developed to contact bulk 
materials \cite{Elec_contacts}. Irrespective of the particularities of the 
material of the contact, the device, or their
geometry, one mainly seeks (i) an Ohmic contact to detect any
non-linearity in the device, and (ii) low resistance to ensure that the
properties measured are those of the device and not those of the contact-device
interface \cite{Hipps}. One of the preferred tools to theoretically simulate 
and extract transport characteristics of low dimensional devices resorts to 
the calculation of non-equilibrium Green's functions (GF) for the system 
composed of the device and the contacts 
\cite{PALee,Meir,Caroli,Datta,Ferry,Jauho}. Its appeal stems from its 
generality and versatility to include in the calculation arbitrary geometries 
of the device, all kinds of external potentials or interactions, electronic 
disorder, etc. Within this framework, in the case of a two-dimensional (2D) 
system the contacts are generally modeled as semi-infinite ballistic ribbons, 
which automatically satisfy the requirement that electrons enter and exit 
the device easily, without returning to it \cite{Ando,LopezSancho,Datta,Ferry}.

Graphene, because of its exceptional mechanical and electronic properties 
\cite{GRoadmap}, has been called to replace existing materials in traditional 
devices such as high frequency and logic transistors \cite{Schwierz_GFET}, 
photodetectors \cite{photodetector}, optical modulators \cite{Opt_modulator},
etc. Its intrinsic two-dimensionality and mechanical robustness is also expected 
to foster a revolution in flexible electronics \cite{flexE}, bio-applications 
\cite{Biosensors}, and energy generation and storage \cite{EStorage}. In current 
and potential applications relying on the electronic degrees of freedom, 
graphene devices must be contacted to a metallic lead. Therefore understanding
how the contact itself impacts the performance of the device is of 
critical importance, both fundamentally and on a more applied level 
\cite{metalGraphene,Barraza-Lopez,edgeeff,Huard}. 

In quantum transport calculations in the context of graphene there are a variety 
of commonly used and accepted models for a contact that meet the requirements 
mentioned above. Common to nearly all these models is the fact that the contact 
geometry eventually converges to a ballistic semi-infinite ribbon at some 
distance from the contact/graphene interface. When the electron dynamics is 
described within the effective Dirac equation approach (i.e. a continuum 
Hamiltonian, rather than a lattice-one) contacts are frequently modeled as 
infinitely doped graphene \cite{infdop}. In the tight-binding model, on the 
other hand, contacts can be either modeled as ideal graphene (hexagonal 
lattice) \cite{Lewenkopf_resumo}, as an ideal metal (square lattice)  
\cite{SchomerusEfC,Schomerus,Blanter}, or using effective models in which the 
effects of the contacts, are reduced to a constant self-energy value 
\cite{SchomerusEfC,extended_contacts,vietnam}. Given that many of the future and 
most unexpected applications of graphene in electronics will rely on patterning 
graphene or transferring it to arbitrarily shaped substrates \cite{Pattern}, 
these tried and tested models of contacts might not always be applicable 
or correct. 

This paper describes an approach to circumvent the difficulties posed by 
these conventional contact models in more generic device layouts. It proposes a 
strategy towards a generic geometry-independent model for the 
contacts, that are then coupled to the usual geometry-dependent tight-binding 
Hamiltonian for the device. The key assumption is that the contacts inject a 
large number of modes close to the Dirac point so that the transport through 
the devices does not depend critically on the specific details (the precise 
mode structure) of the contacts, and their effect can be very well captured by 
an effective self-energy term within the GF's framework \cite{SchomerusEfC}. 
This allows an expedite, flexible, and geometry-agnostic approach which easily 
allows the consideration of device dimensions encompassing more than 5 million 
atoms, and realistic radial dimensions of a few hundreds of nanometers. The 
flexibility of the method is illustrated with calculations including two 
types of disorder, which do not add any significant computational overhead.

\begin{figure}[t]
  \centerline{\includegraphics[width=0.45\textwidth]{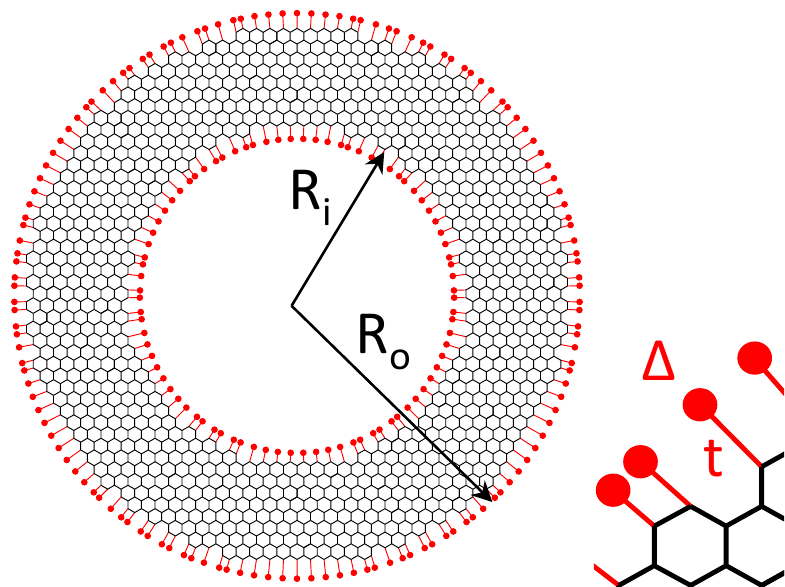}}
  \caption{ (color online ) Left panel: lattice representation of a graphene 
  Corbino disk of inner radius $R_i$ and outer radius $R_o$ contacted to an 
  effective contact . Right panel: a close-up of an edge region showing a 
  schematic of graphene atoms bound to the effective contact. }
  \label{fig:fig1}
\end{figure}

Below we describe this effective-contact approach using the conductance 
calculation in a Corbino geometry as a specific example of its application to 
graphene devices patterned in a non-conventional way. The Corbino disk 
is, in a way, the simplest device for which the contact layout is 
non-trivial, and consists of an annular device region sandwiched between two 
concentric, highly doped, graphene contacts (\Fref{fig:fig1}). Despite its 
importance in the context of understanding the integer quantum Hall effect there 
are only a few studies of ballistic Corbino disks for either Dirac 
\cite{Rycerz_corbino,Rycerz_corbinoB} or Schr\"odinger
\cite{Kirczenow_corbino,Satofumi_corbino,Satofumi_corbino2} electrons. 
Moreover, the circular symmetry, despite a complication for the traditional 
contact models, allows an analytical calculation of the conductance, which we 
will be using as a benchmark for the GF calculation in the ballistic case. 
After thus establishing the robustness of the conductance with respect to 
variations in the number of atoms in the annulus, or at the edges, the 
effective-contact model is used within the GF framework to probe the effect of 
disorder and magnetic fields on the conductance of the Corbino disk. The 
results are in perfect agreement with what is expected from physical grounds, 
as well as related previous calculations on bulk graphene or graphene 
nanoribbons.

\section{Method}

\subsection{Conductance: Dirac equation}

To understand the basic features of the conductance, and to have a conductance 
trace against which to benchmark our results for the lattice model, we outline 
the procedure to extract the conductance using the continuum Dirac description
\cite{Rycerz_corbino}. The single-valley Hamiltonian of graphene can be written 
as $\hat{H}=\hat{H_0} + U\sigma_0$, where
\begin{equation}
\hat{H_0} = -i\hbar v_F\left( \begin{array}{cc}
0 & e^{-i\phi} \{ \partial_r -\frac{i}{r}\partial_\phi \}  \\
e^{i\phi} \{ \partial_r + \frac{i}{r}\partial_\phi \} & 0 \end{array} \right)
.
\label{eq:Hpolar}
\end{equation}
Since $\hat{H}$ commutes with the total angular momentum operator
$\hat{J_z}=-i\hbar \partial_{\phi} + \hbar\sigma_z/2$, the energy eigenstates
with energy $E=\tilde{E}-U$ have the form 
\begin{equation}
  \psi_j(r,\phi)=e^{(j-1/2)\phi}
    \begin{bmatrix}
      \phantom{e^{i\phi}}\chi_{1,j}(r)\\e^{i\phi}\chi_{2,j}(r)
    \end{bmatrix}
    \equiv e^{(j-1/2)\phi}\chi_j(r)
\end{equation}
where $j=\pm1/2,\pm3/2,...$ is the eigenvalue of  $\hbar^{-1}\hat{J_z}$. Without 
loss of generality it is assumed that (i) there is an infinite electron doping 
$E=\tilde{E}-U_\infty > 0$ in the contacts, and (ii) electrons incident from the 
inner contact ($r \le R_i$) are scattered in the Corbino disk ($R_i \le r \le 
R_o$) and finally collected in the outer contact ($R_o \le r $). Under these 
assumptions, the radial component of the wave function in each region is written 
as
\begin{equation}
  \chi_j^i =  \left[ 
  \begin{array}{c} 
  H^{1}_{j-1/2}(K_{\infty} r) \\ 
  iH^{1}_{j+1/2}(K_{\infty}r)e^{i \phi}
  \end{array} 
  \right]
  +
  r_j \left[
  \begin{array}{c} 
  H^{2}_{j-1/2}(K_{\infty}r) \\ 
  iH^{2}_{j+1/2}(K_{\infty}r)e^{i \phi}
  \end{array} 
  \right],
  \label{eq:psi_in}
\end{equation}
\begin{equation}
  \chi_j^c = a_j \left[
  \begin{array}{c} 
  H^{1}_{j-1/2}(kr) \\ 
  iH^{1}_{j+1/2}(kr)e^{i \phi}
  \end{array} 
  \right] 
  +
  b_j \left[ 
  \begin{array}{c} 
  H^{2}_{j-1/2}(kr) \\ 
  iH^{2}_{j+1/2}(kr)e^{i \phi}
  \end{array} 
  \right],
  \label{eq:psi_disco}
\end{equation}
\begin{equation}
  \chi_j^o = t_j \left[
  \begin{array}{c} 
  H^{1}_{j-1/2}(K_{\infty}r) \\ 
  iH^{1}_{j+1/2}(K_{\infty}r)e^{i \phi}
  \end{array} 
  \right],
  \label{eq:psi_out}
\end{equation}
where $H_n^{1(2)}(kr)$ is the Hankel function of the first(second) kind and
$k=E/(\hbar v_F)$; for the highly doped contacts 
$K_\infty=(\tilde{E}-U_\infty)/(\hbar v_F)$ with $U_\infty \rightarrow 
-\infty$. The transmission $t_j$ and reflection $r_j$ amplitudes for each 
channel are obtained by the matching conditions $\psi^i_j(R_i)=\psi^c_j(R_i)$ 
and $\psi^c_j(R_o)=\psi^o_j(R_o)$. Introducing the transmission probability per 
angular momentum channel, $T_j=t_jt_j^*$, the conductance of a graphene Corbino 
disk (including the valley degeneracy) reads 
\cite{Rycerz_corbino,Kirczenow_corbino}
\begin{equation}
G=\frac{4e^2}{h} \sum_j T_j
.
\label{eq:Gdirac}
\end{equation}
This expression allows a direct computation of the conductance as a function 
of the Fermi energy. When the conductance obtained from \Eqref{eq:Gdirac} is 
compared with the GF calculation on the lattice the dimensionless radial 
coordinate $kr$ is related to the tight-binding hopping parameter $t$ and the 
carbon-carbon distance $a=0.142$\,nm via $kr=\frac{2}{3} \left( \frac{E}{t} 
\right) \frac{r}{a}$.

\subsection{Conductance: Lattice Green's Functions}

The starting point of any conductance calculation for non-interacting
electrons using GF is Caroli's formula \cite{Caroli,Datta,Meir}:
\begin{equation}
G=\frac{2e^2}{h} \text{Tr}[\Gamma_qG^r\Gamma_pG^a]
.
\label{eq:Gcaroli}
\end{equation}
Here $G^r=[G^a]^{\dagger}=[E+i\eta-H-\Sigma_p-\Sigma_q]^{-1}$ is the retarded 
GF, $\Gamma_q=i[\Sigma_q-\Sigma_q^{\dagger}]$ reflects the coupling between the 
contacts and the device, and $\Sigma_q$ is the self-energy of contact $q$. 
Eq.~\ref{eq:Gcaroli} has been extensively used in ribbon geometries where one 
contact is to the left and the other to the right of the device region 
\cite{Datta,Lewenkopf_resumo}. The device region can have different shapes or 
can be connected with more than two contacts but, however, each contact has been 
almost invariably modeled as a semi-infinite ribbon 
\cite{UBaranger,GuineaCMap,yo4}. In a Corbino disk an annulus-shaped device 
region is located between two concentric metallic contacts. Eq.~\ref{eq:Gcaroli} 
holds for any contact layout, and device pattern, provided that the contacts and 
device are assumed to have been disconnected in the far past \cite{Meir,Jauho}. 
The total Hamiltonian is then expressed in terms of three contributions 
$$\hat{H}=\hat{H}_{q}+\hat{H_d}+\hat{H}_T,$$ which are the contact Hamiltonian 
$$\hat{H}_{q}=\sum_{k\alpha}\epsilon_{k\alpha}c_{k\alpha}^{\dagger}c_{k\alpha},
$$ the device or central Hamiltonian 
$$\hat{H_d}=\sum_n\epsilon_nd_n^{\dagger}d_n + U(d_n^{\dagger},d_n),$$ where 
$U(d_n^{\dagger},d_n)$ is a one-body potential, and  the contact-device 
tunneling Hamiltonian $$\hat{H}_T=\sum_{k,\alpha,n}  
\left[ V_{k\alpha,n}c_{k\alpha}^{\dagger}d_{n} + h.c.\right].$$
Combining this with the expression for the current from contact $q$, 
$J_q=-e\langle \dot{N_q}\rangle=-ie/\hbar \left[\hat{H},\hat{N_q}\right]$, one 
obtains \eqref{eq:Gcaroli} \cite{Meir,Jauho}. The geometry of the device region 
is easily included in the $\hat{H_d}$ term and the geometry of the contacts is 
encoded in the self energy term. In a tight-binding representation the latter is 
simply expressed as $$\Sigma_q=V_{dq}g_qV_{qd},$$ where $V_{dq(qd)}$ are the 
hopping matrices between the device and the contact, and $g_q$ is the GF of the 
isolated contact. In a generic situation where the geometry of the contacts is 
not amenable to modeling as a semi-infinite ribbon, or any other simple geometry 
allowing an analytical form, obtaining $g_q$ and its tight-binding 
representation would, in principle, be the most challenging step.
Irrespective of the geometry or model used for the contacts the self-energy is a 
complex function, $\Sigma_q=\Lambda -i\Delta$ \cite{Datta}, the real part
describing the shift of the energy levels in the device, and the imaginary part
the broadening of those same levels. Assuming that the contact has an 
approximately constant DOS around the Fermi energy of the device, and that the 
contact only affects atoms at the edge, the self-energy can be simplified to a 
diagonal form
\begin{equation}
\Sigma_q=-i\Delta=-i\pi \rho_c |t_{dq}|^2
,\label{eq:autoE}
\end{equation}
where $\rho_c$ is the contact DOS per atom at the Fermi level
\cite{Datta,Datta_QT_molecule}, and $t_{dq}$ is the coupling between the contact
and the device. 

Transport calculations in graphene are expected to be insensitive to the contact 
model used, provided that the contacts inject a large number of modes close to 
the Dirac point \cite{SchomerusEfC}. One way to guarantee that is to model 
$\rho_c=\rho_\text{graphene}(E=t)=2 t /(\sqrt{3}\pi t^2)=2 /(\sqrt{3}\pi t)$, 
with $t$ being the graphene tight-binding hopping amplitude. In the 
tight-binding problem, the highest DOS occurs precisely at $E=t$, where the 
spectrum exhibits a van~Hove singularity. Hence, in order to mimic a highly 
doped graphene contact, it is natural to set its Fermi level at the van~Hove 
singularity. Under these assumptions the self-energy term can then be set to 
\begin{equation}
\Sigma_q=-i \bigl( 2/\sqrt{3} \bigr) t \approx -it 
,\label{eq:Sigma-effective}
\end{equation}
where, in addition to replacing $\rho_c$ as described above in \Eqref{eq:autoE}, 
we assumed a smooth junction between the contacts and the device: $t_{dq}=t$. 
For the definite case of the Corbino disk that we shall be concentrating on, 
the Hamiltonian of the device (annulus) consists of a nearest-neighbor uniform 
tight-binding approximation. The system is finite, its extent at the 
microscopic level being determined by the condition $R_i \le r \le R_o$ for 
the distance, $r$, of an atom to the origin (see \Fref{fig:fig1}). The 
local, on-site, energy of the atoms with only two neighbors (located at the 
inner and outer edges) is modified by the self-energy term $-it$, and the 
device conductance is calculated using \Eqref{eq:Gcaroli}.

We note that, as an alternative to a conductance calculation in the actual 
honeycomb lattice, the circular symmetry suggests a polar grid discretization 
of the wave equation \cite{Modular_GF}, which could in principle afford an 
opportunity to simulate ideal metallic contacts. As an illustration, the
contact 
GF of an effective annular metallic contact is presented in 
Appendix~\ref{sec:appendix}. Unfortunately, the conductance obtained by this 
scheme is highly sensitive to the contact and coupling parameters because a 
commensurate lattice discretization in the contacts is not possible due to 
the irregularity of the edges in the graphene annulus (see \Fref{fig:fig1}). 
This fact restricts the proper injection of contact modes across the junction 
and, as underlined in the introductory section, the calculated conductance 
becomes dominated by the properties of this junction, rather than by the 
intrinsic behavior of the target annulus region.

\section{Dirac equation \textit{vs} Green's Functions}

\begin{figure}[t]
\centerline{\includegraphics[width=0.45\textwidth]{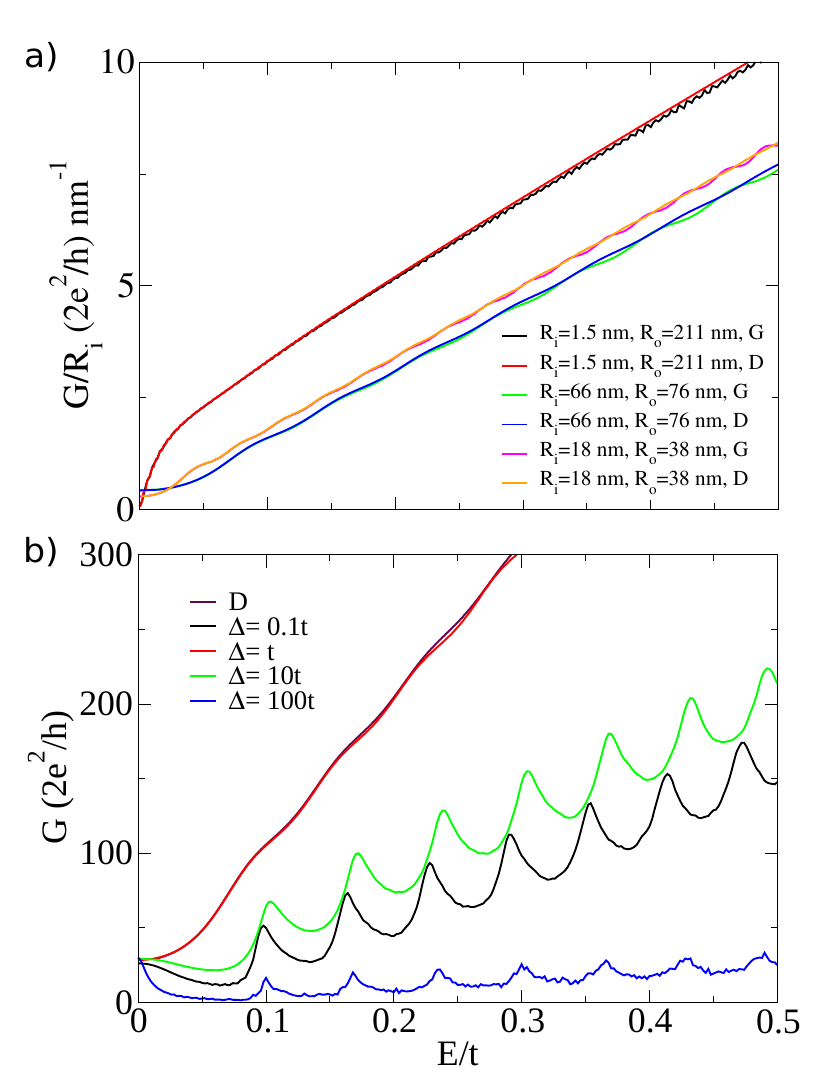}}
\caption{ (Color online) (a) Energy dependence of the conductance normalized by
$R_i$ for different values of $R_i$ and $R_o$, calculated using 
\Eqref{eq:Gdirac} (labeled ``D'') and the lattice Green's function approach 
(labeled ``G''). (b) Conductance of a Corbino Disk with $R_i=66$ nm and $R_o=76$ 
nm using the Green function approach with the effective self-energy term 
$\Sigma_q=-i\Delta$, for different values of $\Delta$. The line labeled by ``D'' 
was calculated using \Eqref{eq:Gdirac} }
 \label{fig:compG}
\end{figure}

To easily compare the values of conductance in Corbino disks of different
aspect ratio $R_i/R_o$ calculated via the Dirac equation \Eqref{eq:Gdirac} and 
the lattice GFs \Eqref{eq:Gcaroli} we normalized the conductance by $R_i$ 
\cite{Rycerz_corbino}.  The meaning of this normalization procedure will be
addressed in the next section. In \Fref{fig:compG}a the normalized conductance 
is shown as a function of the Fermi energy in the annulus region for
$R_i/R_o=0.007,0.86,0.47$. This range of geometric parameters was chosen 
to analyze the effect of the disk size (number of atoms) on the conductance, 
and also to allow us to probe the effect of varying the number of edge atoms 
(which, as advanced above, have their on-site energy modified by the 
self-energy term). These three aspect ratios are achieved in practice with 
three devices with the following characteristics, in order: (i) a wide annulus 
defined by $R_i=1.5$\,nm and $R_o=211$\,nm, with a very large number of carbon 
atoms ($N=5,337,206$), and a large unbalance between the number of atoms on the 
inner ($n_i=42$) and outer edges ($n_o=5,940$); (ii) a narrow annulus with 
$R_i=66$\,nm and $R_o=76$\,nm made out of $N=170,046$ atoms, and having a 
similar number of edge atoms ($n_i=1,860$, $n_o=2,142$); (iii) an 
intermediate case with $R_i=18$\,nm and $R_o=38$\,nm having $N=134,631$ 
atoms, and with a $n_i/n_o$ ratio of about half ($n_i=510$, $n_o=1,071$). 
From \Fref{fig:compG}a it is evident that the conductance calculated using GF 
($G_G$) agrees with the conductance obtained via the Dirac equation ($G_D$), 
irrespective of the geometric parameters. For all geometries both methods 
lead to conductance traces that are hardly distinguishable in the 
plot. Indeed, for the whole energy range shown in this figure, the 
relative difference between the two calculation methods is smaller than 2\,\%, 
confirming that taking $\Delta=t$ produces a smooth effective coupling 
between the highly doped graphene contacts and the graphene annulus. 

In simplified models the contact can be treated as a quantum wire in the wide 
band limit ($t \rightarrow \infty$), which leads to an imaginary constant 
self-energy term ($-i\Delta$). The conductance calculated in that framework 
oscillates when the number of atoms in the device ($N$), the number of atoms on 
the edges ($n_{i(o)}$), or the value of $\Delta$ are changed 
\cite{mujica_QW,extended_contacts}. \Fref{fig:compG}a shows  
oscillations in $G_{D(G)}/R_i$, which are more pronounced
in the GF method due to the roughness of the edges, the scattering at the edges 
is also responsible for the slightly lower value of $G_G$. The period of the 
oscillations is related with the annulus' width $W=R_o-R_i$ 
\cite{Rycerz_corbino}, since it is also observed in $G_D$ and doesn't 
depends on the values of $N$, $n_i$ or $n_o$. The conductance of
Corbino Disks calculated by the GF method behaves similarly to the simplified
wide-band model under changes in $\Delta$ \cite{mujica_QW,extended_contacts}, as
can be seen in \Fref{fig:compG}b. Increasing $\Delta$ improves
transmission until reaching the optimum --compared with the line labeled ``D'', 
calculated using \Eqref{eq:Gdirac}-- conductance line shape for $\Delta=t$.  
Larger $\Delta$ values reduce the conductance again until the peaks 
corresponding to different eigenstates can not be distinguished\cite{Datta}. 

\begin{figure}[t]
\centerline{\includegraphics[width=0.45\textwidth]{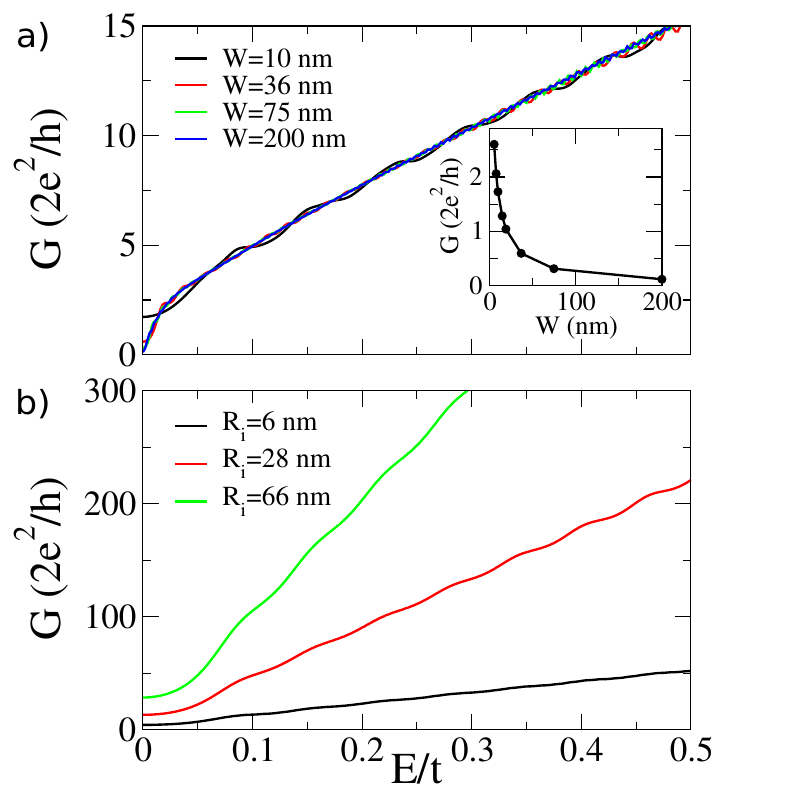}}
\caption{ (Color online) (a) Conductance of a Corbino disk of fixed 
$R_i=1.5$\,nm and varying width $W=R_o-R_i$. The inset shows the
conductance at the Dirac point as function of the width. (b)
Conductance of a Corbino disk of fixed width $W=10$\,nm and
different values of $R_i$ and $R_o$. }
 \label{fig:Gbal}
\end{figure}

\section{Ballistic Conductance of a Corbino Disk}

\subsection{Pristine graphene lattice}

Having shown that the GF method with an effective self-energy term ($\Sigma_q= 
-it$) reproduces the ballistic conductance of graphene Corbino disks we 
now scrutinize its features in more detail. Setting 
first $R_i = 1.5$\,nm and increasing $R_o$, Corbino disks of different width 
are defined. Their conductance characteristic is presented in  \Fref{fig:Gbal}a.
One observes that, firstly, the conductance at the Dirac point (inset) is higher 
for narrow disks due to the evanescent states, and when the width increases 
their effect is reduced and the conductance decreases as $\propto 1/W$ 
\cite{Rycerz_corbino}, reaching $G=0.1(2e^2/h)$ for $W = 200$\,nm.
For higher energies conductance plateaus are not defined due to the highly 
doped 
contacts. Secondly, faint Fabry-Perot oscillations slightly modulate the 
curves of $G(E)$ with a periodicity $\Delta E = \pi \hbar v_F/W$. 
Finally, regardless of the value of the outer radius, the conductance increases 
linearly with the Fermi energy. The slope obtained from \Fref{fig:Gbal}a is 
$23.56$ in units of $\left (2e^2E/ht \right)$. This value can be extracted from 
\Eqref{eq:Gdirac} with a semi-classical argument assuming that the propagating 
angular momentum channels are transmitted with probability one, and that, for a 
given energy $E=v_F\hbar k_F$, the maximum angular momentum eigen-mode that can 
propagate is determined by $j_{\text{max}} \sim \hbar k_F R_i$ when $k_F R_i 
\gg 1$. Under these conditions $G \approx (4e^2/h) 2 j_{\text{max}} = (4e^2/h) 
2 ( 2/3a ) R_i (E/t)$ which, for $R_i=1.5$\,nm used in \Fref{fig:Gbal}, means a 
slope of $28.17$. The discrepancy between this estimate and the numerical slope 
simply reflects the fact that the transmission probability is not one for all 
the angular momentum channels, as can be expected from the fact that the 
effective radial potential depends on the angular momentum 
\cite{Kirczenow_corbino}. From the discussion above one should expect  
higher values of conductance for higher inner radii. This is confirmed in 
\Fref{fig:Gbal}b where the conductance is calculated for fixed $W=10$\,nm and 
varying $R_i$. The effect of a higher $R_i$ is not only a larger slope in 
the $G(E)$ traces at high energies, but also a higher conductance at the Dirac 
point by virtue of the fact that larger inner radii support more total angular 
momentum channels, and hence more modes can be injected into the device. Near 
the Dirac point the conductance is linear in $R_i$ and energy independent. This 
occurs in an energy range $\Delta E \sim 3a/4R_i$, and this energy scale can be 
obtained within a semi-classical approximation recalling that near the Dirac 
point only one value of total angular momentum is allowed: $kR_i=1/2$. The 
radial conductivity defined by \cite{Rycerz_corbino,Shikin}
\begin{equation}
  \sigma_{rr}(E)=G(E)\,\left[\frac{1}{2\pi} \log \frac{R_o}{R_i} \right]
\end{equation}
can be seen to collapse on a single curve at the lowest energies, and then 
eventually on a single point at $E=0$, irrespective of the value of $R_i$ for 
narrow annuli ($R_i/Ro \approx 1$). This universal value corresponds to the 
well known universal minimum of conductivity $\sigma_0=4e^2/\pi h \approx 
0.6 \left(2e^2/h\right)$, whereas for high energies the trace of $\sigma_{rr}$ 
fans out with a slope that is geometry dependent.

\begin{figure}[t]
  \centerline{\includegraphics[width=0.48\textwidth]{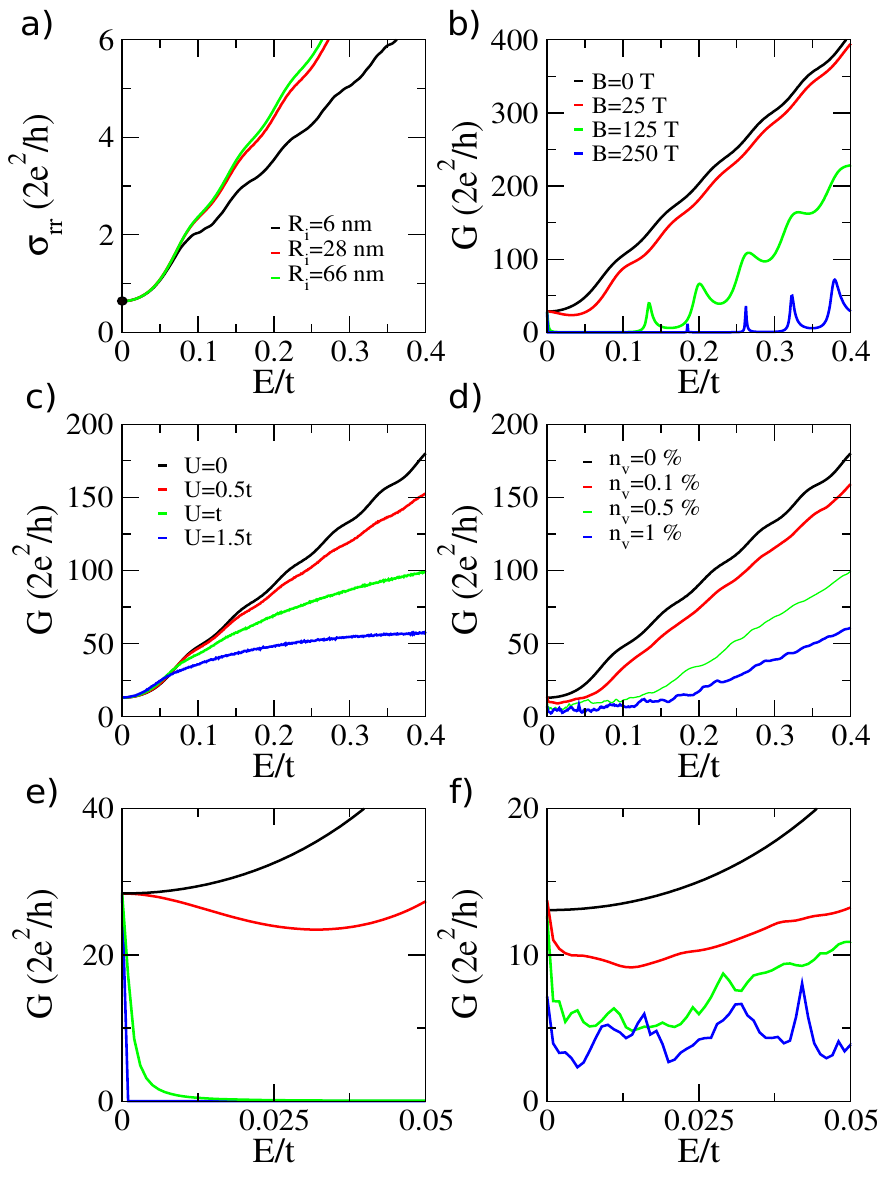} }  
  \caption{ (Color online) (a) Radial Conductivity of a Corbino disk with fixed
  width $W=10$\,nm and different values of $R_i$ and $R_o$. The black dot in 
  the vertical axis marks the value of the universal minimum of conductivity, 
  $\sigma_0=4e^2/\pi h \approx 0.6 \left(2e^2/h\right)$. 
  (b) Conductance of a Corbino disk having $R_i=66$\,nm and $R_o=76$\,nm for 
  different values of an external, perpendicular and homogeneous magnetic 
  field, $B$. The panels in the second row show the 
  conductance of a Corbino disk having $R_i=28$\,nm and $R_o=38$\,nm with 
  on-site disorder (c), and a finite density of vacancies (d). In (e) and (f) 
  we present a close-up of the low energy region for the curves in (b) and 
  (d), respectively. 
  }
  \label{fig:condB}
\end{figure}

In \Fref{fig:condB}b one can see the effect of a constant magnetic field $B$
on the conductance calculated using GFs for a Corbino disk of $R_i=66$\,nm and 
$W=10$\,nm. The conductance line-shape can be straightforwardly understood 
by direct comparison of the electron's cyclotron radius $r_c=\ell_B^2k_F$ 
($\ell_B=\sqrt{\hbar/eB}$ is the magnetic length) with the width ($W$) of the 
disk \cite{Rycerz_corbinoB,Kirczenow_corbino}. As long as $r_c < W/2$ the 
electrons entering from the inner contact cannot reach the outer one, 
which leaves only the possibility of transmission via Landau level (LL) 
assisted resonant tunneling \cite{Prada}. This explains the resonant structure 
of $G(E)$ at precisely the energies $E_n=\left(\hbar v_f/\ell_B\right) 
\sqrt{2n}$, as observed in this figure when $B=250$\,T or $B=125$\,T. For 
example, when $B=250$\,T the maximum energy shown in the plot ($E \approx 
0.4t$) corresponds to $r_c = W/2$, which explains the sharp resonant peaks at  
$E=0,\,0.18t,\,0.26t,\,0.32t,\,E_4=0.37t$. But if $B=125$\,T, resonant peaks 
are observed only at $E=0,\,0.13t$ because at $E \approx 0.2t$ the 
cyclotron radius is already $r_c=W/2$, and hence the third peak at $E=0.18t$ is 
not perfectly defined. When $r_c > W/2$ the injected electrons reach the outer 
contact and the conductance grows linearly in energy. This is seen in all cases 
beyond the particular energy above which that condition holds. Notably, the 
conductance at $E=0$ remains pinned to its zero-field value, irrespective of 
the presence or magnitude of the magnetic field \cite{Prada}. This is shown 
explicitly in \Fref{fig:condB}e where we magnify the low-energy range of the 
curves in \Fref{fig:condB}b: the effect of the magnetic field is to 
sharpen/narrow the conductance peak at $E=0$, without changing its amplitude.

\subsection{Disordered graphene}

The results discussed up to this point have not merely tested the robustness 
and accuracy of the effective self-energy term, but all the results have been 
related and explained on physical grounds. We have seen that the GF method 
perfectly describes the conductance and conductivity of Corbino disks
of different aspect ratios $R_i/R_o$, with and without magnetic field, and for 
a wide range of Fermi energies. The next step is to take advantage of the 
versatility of the GF technique to include disorder. Despite the vast amount 
of studies and attention dedicated to effects of disorder in graphene bulk and 
nano-scale systems \cite{Peres,DasSarma}, the Corbino geometry remains 
unexplored. Since it is well known that the results are affected by the geometry 
of the sample and the model of disorder used, such studies are pertinent and 
relevant, and we proceed now to offer a perspective over some of the 
peculiarities of this problem. The disordered case has been approached for a 
representative system starting with a graphene Corbino disk of $R_i=28$\,nm and 
$R_o=38$\,nm, to which Anderson on-site disorder or a finite density of 
lattice vacancies was added. The extracted conductance is then averaged over 30 
disorder realizations. The results for Anderson disorder are reported in 
\Fref{fig:condB}c, where the on-site energy is uniformly distributed within 
$[ -U/2, U/2]$, and we have considered disorder strengths of $U = 
{0.5t,\,t,\,1.5t}$. At low energies it is clear that the conductance is only 
modestly affected by the on-site disorder. This is due to the fact that for 
$E/t \lesssim 3a/4R_i$ the conductance is dominated by tunneling across the 
entire system via evanescent states, and hence $\sigma_{rr}(E \sim 0) = 
\sigma_0$. The origin of this energy scale can be seen easily with the same 
semi-classical argument used earlier, now applied to the minimum energy above 
which an angular momentum eigen-mode is able to propagate through the system: 
using again the fact that $E=v_F\hbar k_F$, that the angular momentum is 
quantized in half-integer units with $|j|=1/2,3/2,\dots$, and that the 
semi-classical angular momentum corresponds to $j\sim \hbar k_F R_i$, then the 
energy threshold for mode propagation is expected to be $E_{\text{min}} \sim 
j_{\text{min}} v_F / R_i$; replacing $j_{\text{min}}=1/2$ leads to 
$E_{\text{min}}/t = 3a/(4R_i)$. 
At precisely zero energy a conformal mapping can transform a nanoribbon with 
aspect ratio $W/L \gg 1$ into a Corbino disk with $R_i/R_o \approx 1$ 
\cite{Rycerz_corbino,GuineaCMap}. The aspect ratio of the disk studied in 
\Fref{fig:condB}c is 0.86, which means that the radial conductance 
$\sigma_{rr}$ at $E=0$ should coincide with the conductance of a nanoribbon in 
the regime $W \gg L$, which is the so-called pseudo-diffusive regime 
\cite{Rycerz_corbino,Peres,DasSarma}. Our data shows that this value is 
indeed obtained and, since $G(E=0)$ is insensitive to Anderson disorder, the 
corresponding conductivity can be read directly from \Fref{fig:condB}a.
Away from the Dirac point the conformal mapping technique ceases to be valid. 
In this region the conductance decreases as the disorder strength  increases, as 
one expects in general.

In the presence of vacancies or di-vacancies the conductance of Graphene 
nanoribbons exhibits dips, asymmetric Fano resonances, or Breit-Wigner peaks 
\cite{yo2}. This arises due to the fact that a vacancy creates a localized zero 
energy state \cite{Vitorvac}. The interest on the effect of these zero energy 
states and its effect on the electronic transport at the Dirac point has been 
recently boosted  by some experiments \cite{Ponomarenko} and theoretical 
calculations \cite{ANA,Cresti}. Within the tight-binding Hamiltonian that we 
have been considering a vacancy is easily modeled by setting the hopping 
parameter between neighboring atoms to zero, or by setting the on-site energy 
of the vacancy to a value much larger than the energy bandwidth. When a 
single vacancy is included in the Corbino disk there is no effect on the overall
conductance, essentially as a result of the highly doped contacts and the 
radial current distribution. But a finite density of vacancies, $n_v$, equally 
probable on both sublattices \cite{Pereira:2008}, considerably impacts the 
energy dependence of the conductance, as shown in \Fref{fig:condB}d. On the 
one hand, the conductance is peaked at the Dirac point 
with a maximum value that is lower than the conductance of the pristine disk 
and decreases for $n_v > 0.5\%$, while for $n_v \le 0.1\%$ the peak is 5 
\% higher than the pristine conductance, as highlighted in \Fref{fig:condB}f.
This is reminiscent of the formation of the super-metallic regime discussed in 
reference \onlinecite{Cresti}. On the other hand, the conductance exhibits a 
flat energy behavior up to a much larger energy threshold in comparison with 
the behavior in the presence of Anderson disorder, and only then starts growing 
linearly.


\section{Summary}

We established that an effective and simplified model for the contact 
self-energies can be used with high accuracy and robustness in the computation 
of the quantum conductance from lattice Green's functions of nanostructured 
graphene devices with arbitrary contact geometry. As a particular example of 
application, the conductance of graphene Corbino disks was studied in 
detail using this method, the Corbino disk being chosen strategically for being 
a geometry with non-trivial contact configuration for GFs methods, while at the 
same time allowing an analytical solution in the Dirac approximation. This 
permitted the direct comparison and control of the conductance emerging from 
the GF results using the proposed contact model with the conductance that 
follows from the exact solutions in the Dirac approximation. Since many 
envisaged graphene devices and applications entail systems patterned at the 
nano-scale in various configurations that are not always reducible to planar or 
linear contact geometries, an effective and geometry-independent contact model 
such as the one proposed here is certainly a valuable tool for theoretically 
studying the transport characteristics of such structures. Our proposal 
simplifies the description of the contact, but is seen as reliable and, more 
importantly, fulfills the requirements of an ideal contact model, in that it 
does not introduce any spurious features \cite{Hipps}. 
 
As far as the details of the conductance of the graphene Corbino disk are 
concerned, the main results discussed here can be summarized first by 
underlining that the conductance calculated for different aspect ratios 
$R_i/R_o$, with and without magnetic field, shows that one key role of the 
inner contact radius ($R_i$) is to define the maximum number of allowed total 
angular momentum channels and, therefore, the slope of $G(E)$ at high energies. 
The outer radius ($R_o$) is important insofar as it defines the width of the 
annulus, affecting the value of conductance near the Dirac point due to 
tunneling by evanescent states. Finally, we could appreciate how sensitive the 
conductance trace is to the type of electronic disorder: whereas on-site 
Anderson disorder is characterized by the  ``universal'' minimum of 
conductivity ($\sigma_0$) at $E=0$ irrespective of the value of the inner 
radius, strong disorder induced by vacancies, on the other hand, can lead to 
conductance values that exceed the pristine situation near $E=0$, but 
quickly vanish away from the Dirac point.

%
\acknowledgments
We are extremely grateful to Miguel D. Costa for his instrumental collaboration 
in optimizing portions of the numerical computations, and porting some routines 
and operations to a numerically sparse implementation.
This work was supported by the NRF-CRP award ``Novel 2D materials with
tailored properties: beyond graphene'' (R-144-000-295-281).

\appendix

\section{Schr\"odinger Corbino disk}
\label{sec:appendix}

To calculate the conductance of a Corbino disk with Schr\"odinger electrons
using GFs it is necessary to discretize  the Schr\"odinger equation on a polar
grid. Before discretizing, in order to ensure the Hermiticity of the 
Hamiltonian, the transformation $\Psi(r,\phi)=\psi(r,\phi)/\sqrt{r}$ is done 
and the Schr\"odinger equation is re-written as
\begin{equation}
-\frac{\hbar^2}{2m_e} \left[  \frac{\partial^2 }{\partial r^2} + \frac{1}{4r^2}
 + \frac{\partial^2}{\partial \phi^2} \right] \psi + V(r,\phi) \psi =E\psi
.
\label{eq:Sch}
\end{equation}
On the grid the  wave function $\psi(r,\phi)$ is expressed as $\psi_{m,j}$ where
the indexes $(m,j)$ represent the radial, $r=m\Delta_r$, and polar, 
$\phi=j\Delta_{\phi}$, sites, with $\Delta_r$ and $\Delta_{\phi}$ the 
radial and angular grid spacing, respectively. Writing \Eqref{eq:Sch} in a 
finite difference approximation leads to
\begin{multline}
t_r\psi_{m+1,j} + t_r\psi_{m-1,j} \\
- \left[  2t_r + 2t^m_{\phi} +U_m + 
V_{m,j}\right] \psi_{m,j}  + \\ 
t^m_{\phi}\psi_{m,j+1} + t^m_{\phi}\psi_{m,j-1} = E\psi_{m,j} 
,
\label{eq:Sch_fd}
\end{multline}
with the radial hopping, the angular hopping, and the radial potential given, 
respectively, by 
\begin{equation}
  t_r\!=\!-\frac{\hbar^2}{2m_e\Delta^2_r}
  ,\,
  t^m_{\phi}\!=\!-\frac{\hbar^2}{2m_er^2_m\Delta^2_{\phi} }
  ,\,
  U_m\!=\!\frac{\hbar^2}{8m_er^2_m}
  .
\end{equation}

\begin{figure}[h]
\centerline{\includegraphics[width=0.45\textwidth]{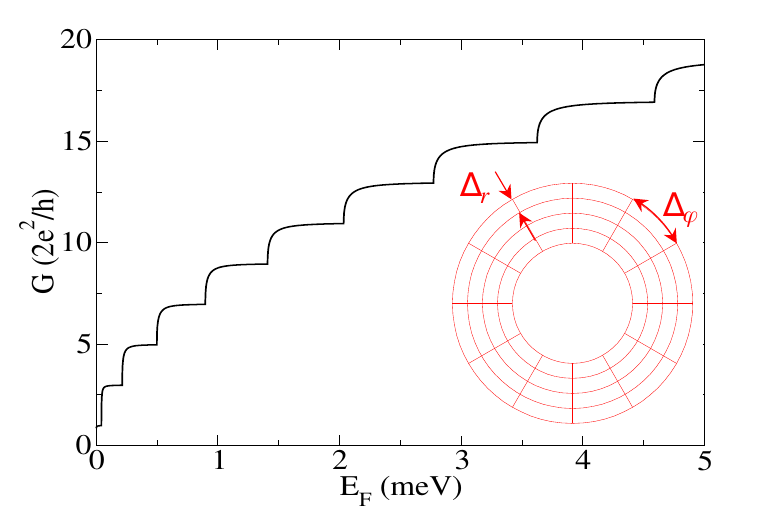}}
\caption{ Conductance of a ballistic Corbino disk ($R_i=0.1$\,$\mu$m and 
$R_o=0.2$\,$\mu$m) in GaAs as a function of the Fermi energy, $E_F$. The curve 
was obtained using the effective contact Green function 
\eqref{eq:GF_contact_sch}. 
Inset: schematic of the polar grid used in the finite difference decomposition 
of the Schr\"odinger equation.
}
\label{fig:GSch}
\end{figure}

In the contacts one assumes that the electrons are under the influence of an 
effective constant potential. This means that one sets 
$U_m=\frac{\hbar^2}{8m_eR^2_{i(o)}}$ and 
$t^m_{\phi}=-\frac{\hbar^2}{2m_eR^2_{i(o)}\Delta^2_{\phi} }$ as constants in 
\Eqref{eq:Sch_fd}. Under these assumptions the GF of the contacts is calculated 
as
\begin{multline}
g(E;R_{i(o)},j,j') = \\
\sum^{N_{\phi}}_{l=0} \left(  \frac{e^{il2\pi 
j/N_{\phi}}}{\sqrt{N_{\phi}}} \right)  \frac{e^{i\theta^l_{i(o)}}}{t_r}  \left(  
\frac{e^{-il2\pi j'/N_{\phi}}}{\sqrt{N_{\phi}}} \right)
\!,
\label{eq:GF_contact_sch}
\end{multline}
where $N_{\phi}$ is the number of cells in the angular grid and
\begin{multline}
\cos\theta^l_{i(o)}=\frac{E + t_r + 2t^{i(o)}_{\phi} + U_{i(o)} } {2t_r}
\\
-\frac{t^{i(o)}_{\phi}}{t_r} \cos\biggl(\frac{2\pi l}{N_{\phi}}\biggr)
\label{eq:valtheta}
.
\end{multline}
\Fref{fig:GSch} shows the result of this approach in the calculation of the 
conductance of a Corbino disk where the target conducting medium is GaAs 
($m_e=0.067m_o$), with $R_i=0.1$\,$\mu$m and $R_o=0.2$\,$\mu$m. 
The main features of the quantum conductance in a \emph{massive} electronic 
system such as this one were previously described by Kirczenow 
\cite{Kirczenow_corbino}. We underline that, despite the fact that one can 
obtain these analytical expressions for the quantum conductance in this case, 
it immediately becomes unpractical in any realistic scenario. First, even though
metallic contacts can be modeled by \Eqref{eq:GF_contact_sch} and coupled to 
a graphene annulus as described above, the conductance trace is highly sensitive 
to the contact parameters, which is not an ideal situation. Secondly, the 
radial decomposition offers little advantage for a geometry that is not 
perfectly cylindrical, which makes the approach of limited use for general 
contact geometries. Finally, even when the perfect Corbino layout is 
considered, the quantization shown in \Fref{fig:GSch} has not been observed 
experimentally, which necessarily raises the question of how to treat disorder 
efficiently within  such an approach. That is when calculations based on lattice 
GFs offer a more flexible and expedite (and, at the same time, less biased, or 
with less approximations) means of extracting the transport quantities. In fact, 
as far as the computational effort of the lattice GFs is concerned, the 
inclusion of disorder has very little detrimental impact.

\bibliographystyle{apsrev}
\bibliography{references}

\end{document}